\documentclass{article}

\usepackage{PRIMEarxiv}

\usepackage[utf8]{inputenc} 
\usepackage[T1]{fontenc}    
\usepackage{hyperref}       
\usepackage{url}            
\usepackage{booktabs}       
\usepackage{amsfonts}       
\usepackage{nicefrac}       
\usepackage{microtype}

\usepackage{lipsum}
\usepackage{fancyhdr}       
\usepackage{graphicx}       
\graphicspath{{media/}}     
\usepackage{setspace}
\usepackage{subcaption}
\usepackage{amsmath}
\usepackage{algpseudocode}
\usepackage{algorithm}
\usepackage{booktabs}
\usepackage{array}
\usepackage[utf8]{inputenc}
\usepackage{tcolorbox}
\usepackage{mdframed}
\definecolor{mycolor}{RGB}{210, 235, 255} 
\definecolor{mycolor2}{RGB}{255, 224, 178}
\usepackage{xcolor}
\usepackage{setspace}
\usepackage{multicol}

\usepackage{subcaption}

\DeclareCaptionFormat{mysub}{#3}  
\usepackage{titlesec}
\titlespacing*{\section}{0pt}{0.3\baselineskip}{0.2\baselineskip}

\title{DualSwinUnet++: An Enhanced Swin-Unet Architecture With Dual Decoders For PTMC Segmentation
}

\author{
  Maryam Dialameh\\
  Department of Mechanical and Mechatronics Engineering\\
  University of Waterloo, Waterloo, ON, Canada\\
  \texttt{maryam.dialameh@uwaterloo.ca} \\
  \And
  Hossein Rajabzadeh \\
  Department of Mechanical and Mechatronics Engineering\\
  University of Waterloo, Waterloo, ON, Canada\\
  \texttt{hossein.rajabzadeh@uwaterloo.ca}
    \And
  Moslem Sadeghi-Goughari \\
  Department of Mechanical and Mechatronics Engineering\\
  University of Waterloo, Waterloo, ON, Canada\\
  \texttt{m8sadegh@uwaterloo.ca}
      \And
  Jung Suk Sim \\
  Department of Radiology \\ Withsim Clinic
Seongnam, Korea \\
  \texttt{jungsuk.sim@gmail.com}
  \And
  Hyock Ju Kwon \\
  Department of Mechanical and Mechatronics Engineering\\
  University of Waterloo, Waterloo, ON, Canada\\
  \texttt{hjkwon@uwaterloo.ca} \\
}

\begin{document}
\maketitle
\begin{abstract}
Precise segmentation of papillary thyroid microcarcinoma (PTMC) during ultrasound-guided radiofrequency ablation (RFA) is critical for effective treatment but remains challenging due to acoustic artifacts, small lesion size, and anatomical variability. In this study, we propose \textbf{DualSwinUnet++}, a dual-decoder transformer-based architecture designed to enhance PTMC segmentation by incorporating thyroid gland context. DualSwinUnet++ employs independent linear projection heads for each decoder and a residual information flow mechanism that passes intermediate features from the first (thyroid) decoder to the second (PTMC) decoder via concatenation and transformation. These design choices allow the model to condition tumor prediction explicitly on gland morphology without shared gradient interference. Trained on a clinical ultrasound dataset with 691 annotated RFA images and evaluated against state-of-the-art models, DualSwinUnet++ achieves superior Dice and Jaccard scores while maintaining sub-200ms inference latency. The results demonstrate the model's suitability for near real-time surgical assistance and its effectiveness in improving segmentation accuracy in challenging PTMC cases.
\end{abstract}.


\section{Introduction}
\label{intro}
Papillary Thyroid Microcarcinoma (PTMC) is characterized by tumors that are 1 cm or smaller in diameter  \cite{dideban2016thyroid}. Despite its small size, PTMC can still pose significant health risks if not managed properly \cite{kaliszewski2020risk}.Radiofrequency Ablation (RFA) has emerged as a minimally invasive treatment option for PTMC treatment, particularly for patients who are not ideal candidates for open surgery \cite{bernardi2021current}. RFA involves the use of high-frequency electrical currents to generate localized heat, effectively destroying cancer cells within the thyroid gland. During the RFA procedure, the physician relies on real-time ultrasound imaging to guide the ablation process \cite{orloff2022radiofrequency}. However, this treatment faces two major challenges: 1) it becomes difficult for the physician to correctly identify the PTMC location and its boundaries; 2) distinguishing between normal tissue and a PTMC can be challenging in many cases \cite{kaliszewski2019papillary}. 

Currently, ultrasound B-mode imaging is employed to guide RFA, allowing physicians to identify the tumor boundaries using images from the targeted area. Although ultrasound imaging offers a quick, cost-effective, and safe imaging solution, the precision of imaging and targeting, as well as the skill required from physicians, remain significant challenges \cite{mohammadkarim2018hemodynamic}. First, ultrasound B-mode imaging suffers from limited resolution, which can lead to low accuracy in identifying tumor boundaries \cite{afrin2023deep}. Additionally, during RFA treatment, each RF sonication generates microbubbles at the targeted area due to the ablation effect. These microbubbles can persist for minutes to hours, causing acoustic shadowing at the boundary of the PTMC. Consequently, subsequent sonications may face difficulties in accurately detecting the tumor boundary \cite{huang2020phase}. This can result in untreated tumor areas or necessitating repeated ablation procedures, increasing the risk of tissue overheating  \cite{shi2019inflammation}. Despite advancements in medical imaging, the task of real-time targeting and tracking of the tumor area during RFA  remains challenging, and clinicians continue to seek effective solutions for these issues.

Artificial Intelligence (AI), particularly deep learning (DL) models, has been successfully employed across various fields, including medical applications \cite{khan2018review}. One of the most prominent uses of deep learning in the medical sector is medical image segmentation  \cite{wang2022medical}. Such technologies are increasingly considered as valuable tools for assisting RFA operators by providing real-time guidance and crucial insights during procedures. Unet, for instance, is a convolutional neural network architecture, and recognized as one of the primer and successful DL methods to tackle the problem of medical image segmentation. Variants of U-net such as V-net \cite{milletari2016v}, U-net++ \cite{zhou2018unet++}, attention U-net \cite{oktay2018attention}, residual U-net \cite{zhang2018road}, and dense U-net \cite{guan2019fully} have been developed to overcome the limitations of the original U-net model, including faster convergence, the ability to handle varied segments sizes, and improving accuracy. 
A recent advantage of Medical SAM Adapter (Med-SAM) \cite{wu2023medical}adapts SAM framework for medical applications. The main idea of Med-SAM is applying low-rank adapters with ReLU activation to bridge the gap between SAM and medical segmentation tasks. Another method, TRFE-Net \cite{gong2021multi}, is an encoder-decode convolutional framework that incorporates prior knowledge of the thyroid region to enhance thyroid nodule segmentation accuracy in ultrasound images. Additionally, vision transformer-based architectures \cite{dosovitskiy2020image,vaswani2017attention} have opened up new possibilities for PTMC segmentation from ultrasound images. For instance, Swin-Unet \cite{cao2022swin} leverages the hierarchical nature of Swin-Transformers \cite{liu2021swin} to handle various scales of features effectively, enhancing performance in segmentation tasks. In another recent study \cite{wang2024predicting}, the authors proposed a deep learning framework combining a ResNet-based deep convolutional neural network (CNN) and an MLP to jointly process ultrasound images and clinical features for central lymph node metastasis prediction in PTMC, demonstrating effective multimodal fusion through a tailored architecture. Multimodal attention-based framework is another work that integrates MRI radiomics and clinical-pathological features using separate encoder networks, followed by channel and spatial attention modules and self-attention fusion to predict central lymph node metastasis (CLNM) in papillary thyroid carcinoma (PTC) \cite{wang2024predicting}; however, the reliance on MRI makes the approach relatively expensive and time-consuming compared to ultrasound-based methods, limiting its real-time clinical utility. More recently, a radiomics-based framework \cite{chen2024ultrasound} was developed using conventional ultrasound imaging, where handcrafted texture, shape, and intensity features were extracted from segmented thyroid nodules and selected through LASSO and cross-validation to train a LightGBM classifier for malignancy prediction, achieving strong diagnostic performance; however, the reliance on manual feature engineering limits adaptability across imaging settings and makes it less robust compared to end-to-end deep learning models.

Challenges in medical images, such as low pixel-intensity variations, class imbalance, organ interference and move- ment, and limited annotated samples, make this area of research particularly complex \cite{hesamian2019deep}. Moreover, current PTMC segmentation solutions for PTMC segmentation during RFA surgery face additional challenges, including the need for real-time prediction, robustness against organ movements in the throat region (as patients only receive regional anesthesia), and the ability to track PTMC amid bubbles created by heat during the procedure \cite{zhou2015monitoring}. Another limitation of current state-of-the-art models is their inability to incorporate thyroid information into their predictions when dealing with PTMC segmentation using ultrasound images. For instance, we experimentally show that Swin-Unet exhibits a significant false positive rate in PTMC segmentation. Moreover, the information flow between a model’s layers play an important role in segmentation accuracies. 

To address the above-mentioned limitations, we propose a novel segmentation model, named DualSwinUnet++, which is an end-to-end learning procedure built on the Swin-Unet model. More specifically, DualSwinUnet++ enhances segmentation accuracy by expanding the Swin-Unet model to incorporate two parallel decoders. The first decoder is designed to integrate the thyroid gland information, which is fed into the second decoder, leading to more precise segmentation results. By restricting the examination of PTMC to within the thyroid gland, we achieve two key benefits. Firstly, we minimize the risk of false positives by ensuring that predictions are confined to the thyroid gland. Secondly, we reduce the model’s search space, which allows it to focus on a smaller, more relevant region. This restriction not only simplifies the learning process but also results in a model with fewer parameters, thereby enhancing prediction speed during inference. Therefore, our key contributions are as follows:
(1) We introduce a dual-decoder architecture where the first decoder segments the thyroid gland and informs the second decoder for PTMC segmentation through a dedicated residual information flow path.
(2) Unlike prior dual-decoder models, we employ \textbf{independent linear projection heads} for each decoder, allowing disentangled gradient optimization that improves learning stability and accuracy.
(3) We design a \textbf{custom residual fusion mechanism} that passes intermediate features from the thyroid decoder to the PTMC decoder via concatenation and linear transformation, enhancing spatial awareness and reducing false positives.
(4) We conduct extensive experiments on two ultrasound datasets, demonstrating that our method achieves state-of-the-art accuracy while maintaining real-time inference speed, making it suitable for AI-assisted RFA interventions.

\section{Proposed Method: DualSwinUnet++}
\begin{figure}[t]
    \centering
    \subfloat[ (a) ]{
        \includegraphics[width=0.7\textwidth]{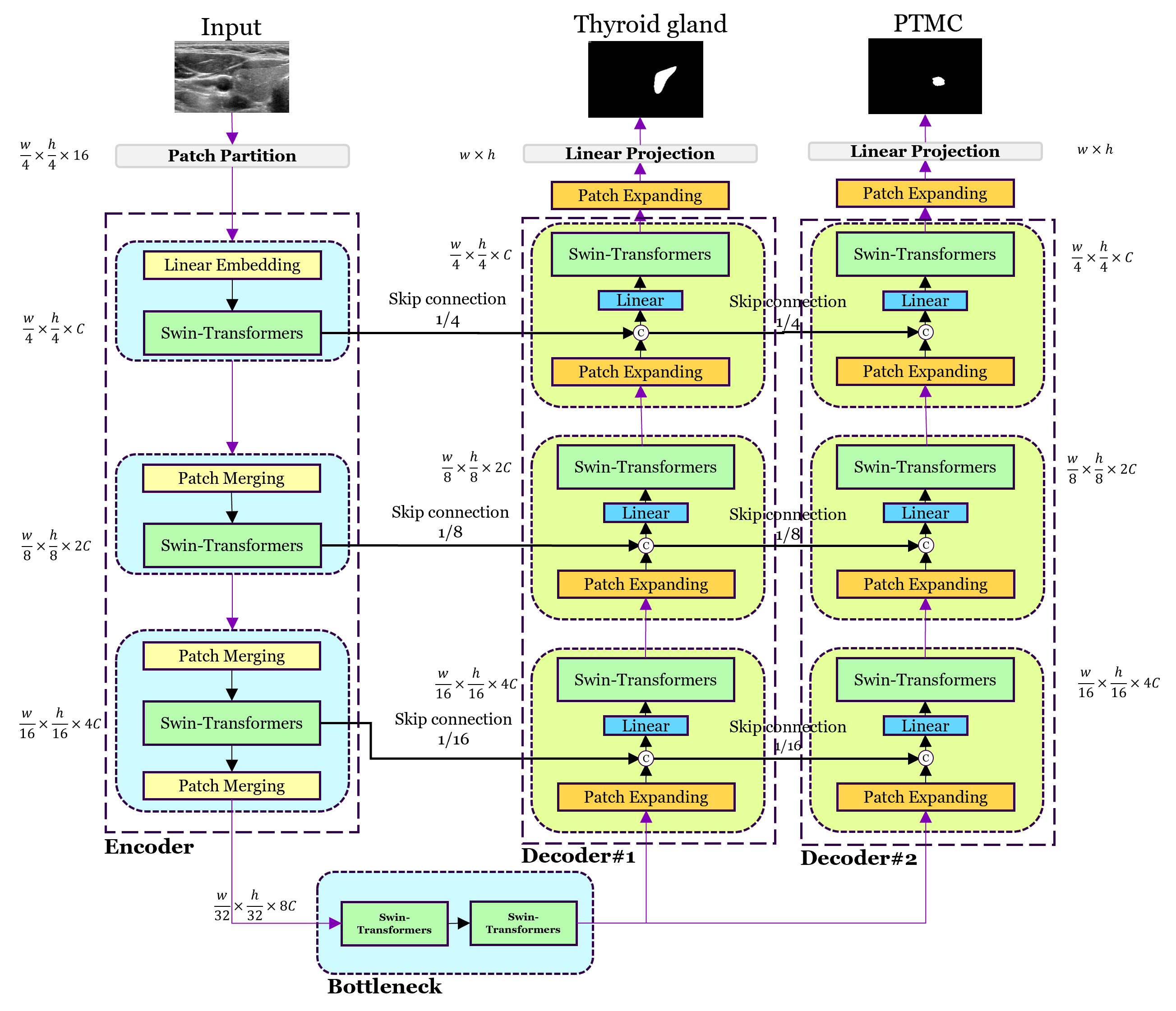}
        \label{fig:arch}
    }\\[1.5ex]
    \subfloat[ (b) ]{
        \includegraphics[width=0.2\textwidth]{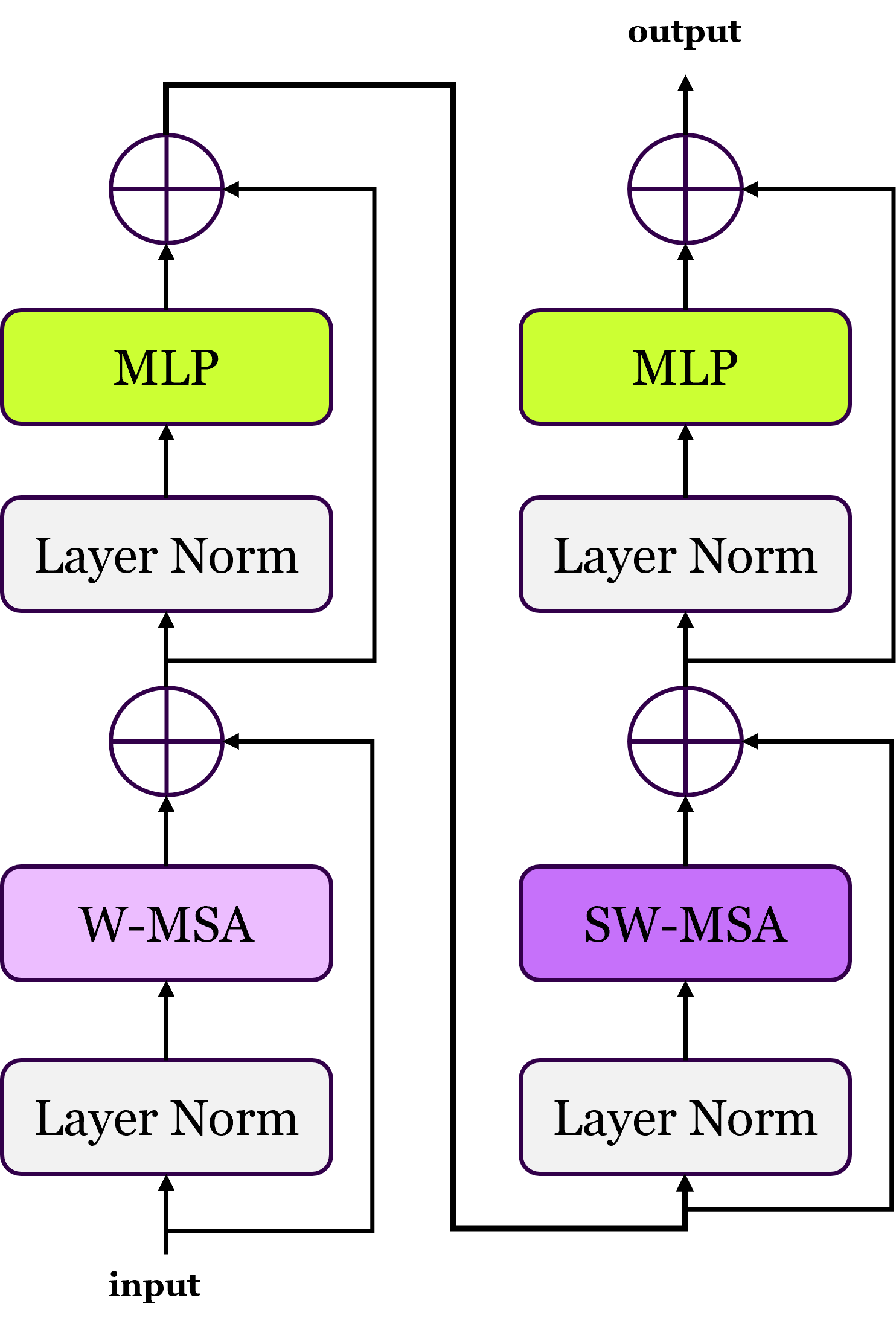}
        \label{fig:swin_block}
    }
    \caption{Overview of the proposed model architecture and its key components. (a) The overall architecture of DualSwinUnet++, consisting of a shared encoder and two parallel decoders.The first decoder segments the thyroid gland, while the second decoder focuses on PTMC segmentation, conditioned on features from both the encoder and the first decoder via skip connections. (b) Swin-Transformer block. Each block contains two stages: a Window-based Multi-Head Self Attention (W-MSA) block and a Shifted Window MSA (SW-MSA) block. Each attention unit is followed by Layer Normalization and a Multi-Layer Perceptron (MLP) module with residual connections, enabling efficient local-global context modeling at each scale.}
    \label{f1}
\end{figure}

This section outlines our proposed model, DualSwinUnet++, designed to address the task of PTMC segmentation from ultrasound images captured during RFA surgery. As illustrated in Figure \ref{f1} (a), the proposed DualSwinUnet++ comprises one encoder and two decoders, each formed by stacking Swin-Unet blocks. The encoder receives the input grayscale patches of an ultrasound image captured during RFA surgery. These patches are first processed through a linear embedding layer, which transforms the raw pixel intensities into a sequence of tokens by splitting the image into $4\times4$ patches without overlap. Hence,each token at this stage is represented by a raw-intensity vector of size $4\times4\times1=16$.  Subsequently, these patches are passed through a second linear embedding layer, converting raw-pixel patches to the embedding space of size $C$. The embedded tokens are passed through three pairs of Swin-Transformers blocks and patch merging layers, with the same functionality as the reference encoder in vanilla Swin-Unet \cite{cao2022swin}. The workflow of the Swin-Transformers block, illustrated in Figure \ref{f1}-(b), consist of two successive self-attention sub-blocks. The first sub-block employs a windowed multi-head self-attention (W-MSA) along with layer normalization and GeLU-activated MLP modules, while the second sub-block uses a shifted-window multi-head self-attention (SW-MSA).

The patch merging layers first normalize its input features and then reduce the spatial resolution by a factor of 2 while increasing the feature embedding size by a factor of 2. This process is computed through a linear projection operation. The encoder outputs are then passed to a bottleneck layer, comprising two consecutive Swin-Transformers layers.

The outputs of the bottleneck are directed towards two decoders: the first decoder predicts the thyroid gland segment, and the second decoder predicts the PTMC segment. Each decoder contains three layers, with each layer comprising a patch expanding module to up-sample the feature vector, a concatenation operator (depicted by \textcopyright) to  concatenate the patch expanding outputs with skip connections, a linear projection module, and finally a Swin-Transformers block. In both decoders, the linear projection modules are used to aggregate the concatenated features and maintain a consistent dimensions. Finally, a linear layer projects the decoders outputs to the generate pixel-level segmentation predictions.

While both decoders are similar to each other with the same architecture, they are different in terms of their input skip connections. Specifically, the first decoder receives the encoder’s intermediate features via skip connections. In contrast, the second decoder has access to information provided by the concatenated intermediate features of both the encoder and the first decoder. Therefore, the second decoder also has access to the information related to the thyroid gland, which further allows the second decoder to make more accurate predictions.

The proposed DualSwinUnet++ is trained using an end-to-end strategy. The training loss function consists of two components: 1) thyroid gland segmentation loss, and 2) PTMC segmentation loss. The first loss computes the prediction error for thyroid segmentation, while the second loss calculates the prediction error for PTMC segmentation. Equation \ref{e1} represents the training loss function, where $\alpha$ and $\beta$ are hyper-parameters,  $DL(.)$ denotes Dice loss, and $BCE(.)$ is the binary cross-entropy loss. Additionally, $X$ represents the set of input samples, $F_1(X)$, $F_2(X)$ are the outputs of the first and second decoders, and $Y_{PTMC}$, $Y_{Thyroid}$ stand for the true labels for PTMCs and thyroids, respectively, associated with the inputs $X$.

Equation \ref{e2} and \ref{e3} respectively describe the BCE and DL losses, where $y_i$ and $\hat{y_i}$ are true label and the predicted probability, and $N$ is the total number of samples.

\section{Experiments}
\label{sec:Experiments}
\begin{figure}[ht]
    \centering
    \begin{subfigure}[b]{0.3\textwidth}
        \centering
        \includegraphics[width=\textwidth]{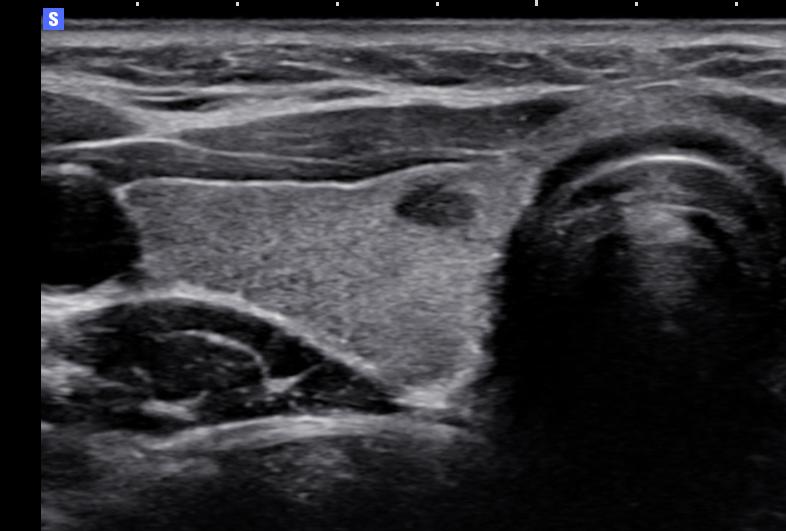}
        \caption{Sample image 1}
        \label{fig:image1}
    \end{subfigure}
    \begin{subfigure}[b]{0.3\textwidth}
        \centering
        \includegraphics[width=\textwidth]{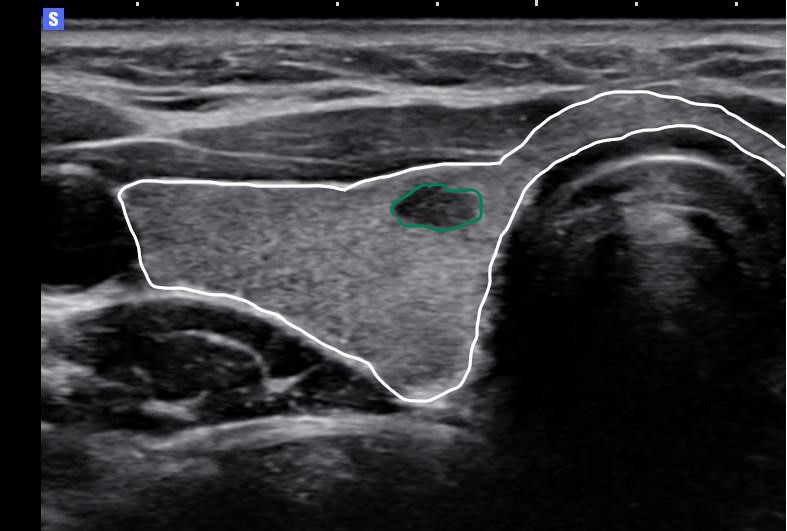}
        \caption{Sample image 2}
        \label{fig:image2}
    \end{subfigure}
    
    \vspace{0.3cm} 

    \begin{subfigure}[b]{0.3\textwidth}
        \centering
        \includegraphics[width=\textwidth]{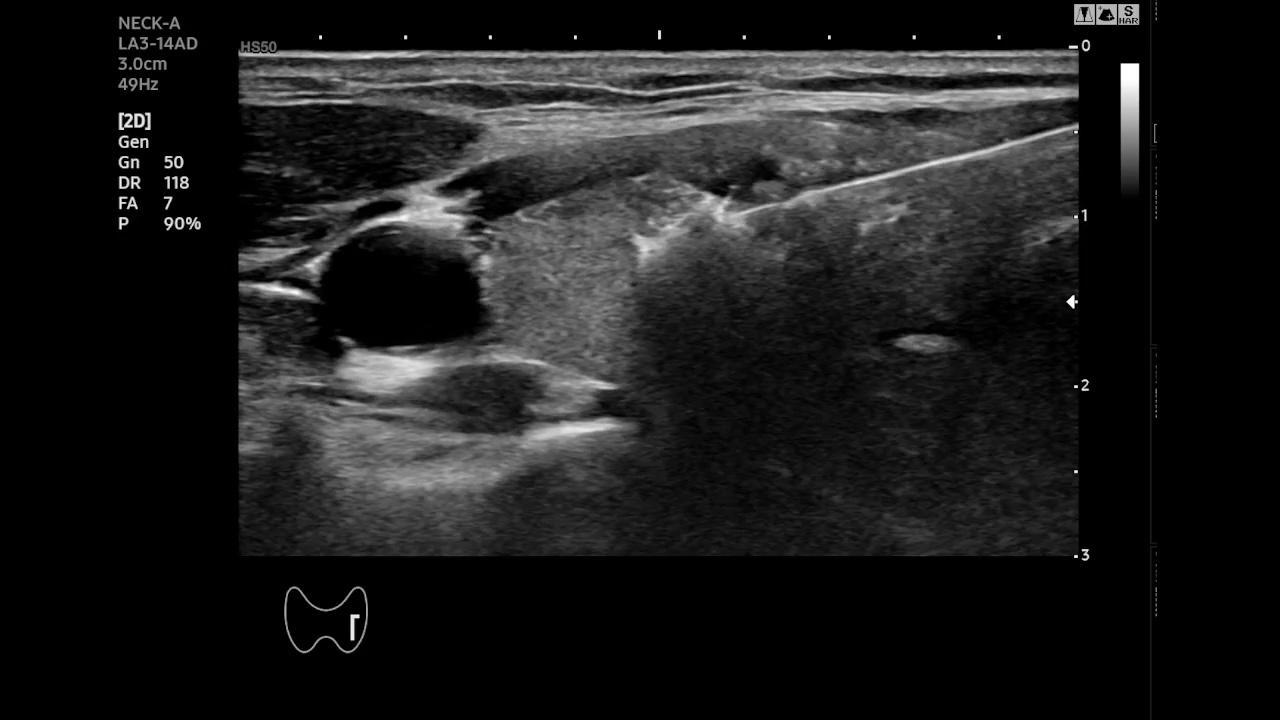}
        \caption{Sample image 3}
        \label{}
    \end{subfigure}
    \begin{subfigure}[b]{0.3\textwidth}
        \centering
        \includegraphics[width=\textwidth]{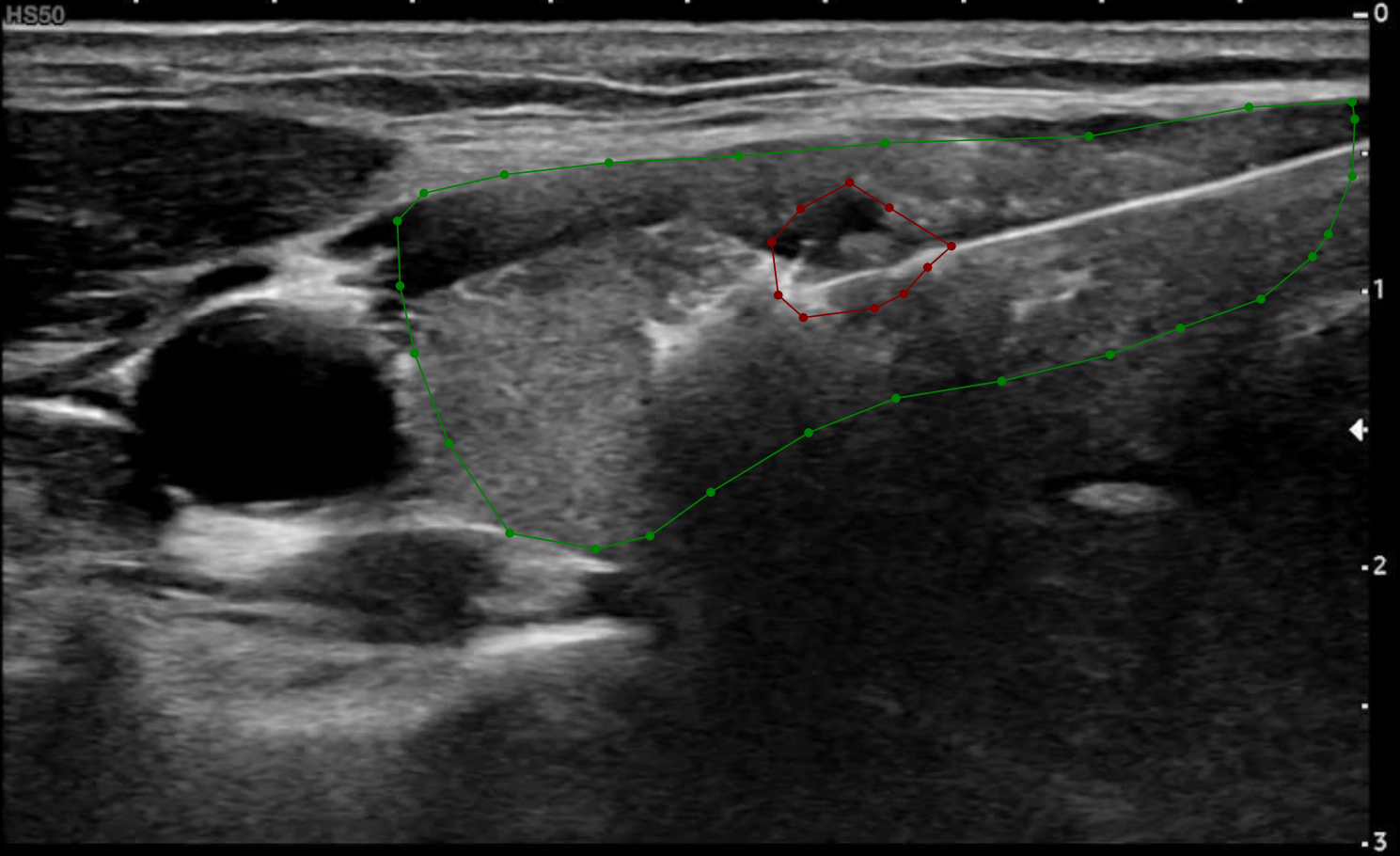}
        \caption{Sample image 4}
         
    \end{subfigure}

    \caption{Sample ultrasound images from the dataset used in the experiments, along with their corresponding annotations. The images represent different instances of PTMC (Papillary Thyroid Microcarcinoma), highlighting the variability in appearance and the complexity of segmentation tasks. }
    \label{f2}
\end{figure}

This section presents the evaluation and comparison results of our proposed DualSwinUnet++ model against several relevant models. We utilized two datasets for the evaluation. The first dataset consists of 691 annotated ultrasound images collected from 101 anonymous patients who underwent RFA surgery. The dataset is divided into training, validation and test sets with proportions of $80 \%, 10 \%, 10 \%$, respectively. The images are in jpg-format with a resolution of $786 \times 531$ pixels. We annotated them in collaboration with specialists, and all annotations were approved under their supervision. Figure \ref{f2} show sampled images from our dataset alongside their associated annotations.

The second dataset used for evaluation is TN3k \cite{gong2021multi}, consisting of thyroid nodule images along with their associated masks. TN3K includes 3,493 ultrasound images collected from 2,421 cases.  
In the context of image segmentation, two commonly used evaluation metrics are the Jaccard Index and the Dice Coefficient \cite{muller2022towards}. The Jaccard Index, also known as the Intersection over Union (IoU), is defined as the size of the intersection divided by the size of the union of the sample sets. Mathematically, it can be expressed as:

\begin{equation}
J(A, B) = \frac{|A \cap B|}{|A \cup B|}
\end{equation}

where \(A\) and \(B\) present the ground truth and the predicted segmentation, respectively. The Jaccard Index ranges from 0 to 1, where 0 indicates no overlap and 1 indicates perfect overlap. The Dice Coefficient, also known as the Sørensen–Dice Index, is a statistic used to gauge the similarity between two samples. It is calculated as twice the size of the intersection divided by the sum of the sizes of the two sets. The formula for the Dice Coefficient is:

\begin{equation}
D(A, B) = \frac{2|A \cap B|}{|A| + |B|}
\end{equation}

\begin{equation}
\begin{aligned}
     &Loss(X,Y_{Thyroid}, Y_{PTMC}) = \\
     &\frac{1}{2}\Big( \alpha \cdot BCE\big( F_1 (X), Y_{Thyroid} \big) + (1-\alpha) \cdot DL\big( F_1 (X), Y_{Thyroid} \big) + \\ 
     &\beta \cdot BCE\big( F_2(X), Y_{PTMC} \big) + (1-\beta) \cdot DL\big( F_2 (X), Y_{PTMC} \big) \Big)
\end{aligned}
    \label{e1}
\end{equation}

\begin{equation}
\text{BCE}(y, \hat{y}) = -\frac{1}{N} \sum_{i=1}^{N} \Big( y_i \cdot \log(\hat{y}_i) + (1 - y_i) \cdot \log(1 - \hat{y}_i) \Big)
\label{e2}
\end{equation}

\begin{equation}
\text{Dice Loss}(y, \hat{y}) = 1 - \frac{2 \sum_{i=1}^{N} y_i \hat{y}_i}{\sum_{i=1}^{N} y_i + \sum_{i=1}^{N} \hat{y}_i}
\label{e3}
\end{equation}



\begin{figure}[ht]
    \centering
    \begin{subfigure}[b]{0.15\textwidth}
        \centering
        \includegraphics[width=\textwidth]{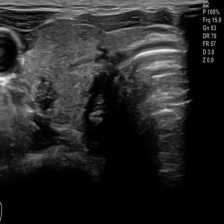}
        \caption{Image 1}
         
    \end{subfigure}
    \begin{subfigure}[b]{0.15\textwidth}
        \centering
        \includegraphics[width=\textwidth]{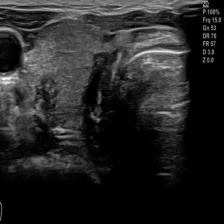}
        \caption{Image 2}
        \label{fig:image6}
    \end{subfigure}
    \begin{subfigure}[b]{0.15\textwidth}
        \centering
        \includegraphics[width=\textwidth]{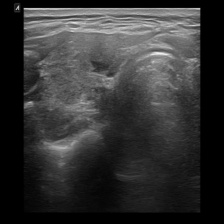}
        \caption{Image 3}
        \label{fig:image7}
    \end{subfigure}
    \begin{subfigure}[b]{0.15\textwidth}
        \centering
        \includegraphics[width=\textwidth]{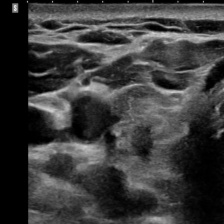}
        \caption{Image 4}
        \label{fig:image8}
    \end{subfigure}
    \begin{subfigure}[b]{0.15\textwidth}
        \centering
        \includegraphics[width=\textwidth]{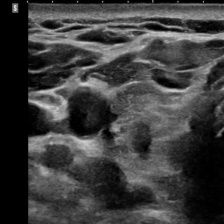}
        \caption{Image 5}
        \label{fig:image9}
    \end{subfigure}
    \begin{subfigure}[b]{0.15\textwidth}
        \centering
        \includegraphics[width=\textwidth]{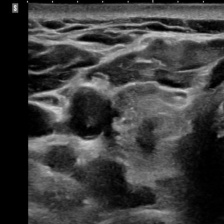}
        \caption{Image 6}
        \label{fig:image10}
    \end{subfigure}
    \vspace{0.15cm} 

    \begin{subfigure}[b]{0.15\textwidth}
        \centering
        \includegraphics[width=\textwidth]{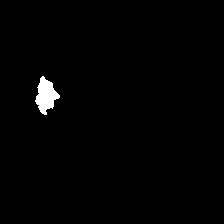}
        \caption{Actual 1}
        \label{}
    \end{subfigure}
    \begin{subfigure}[b]{0.15\textwidth}
        \centering
        \includegraphics[width=\textwidth]{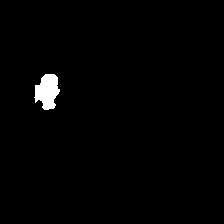}
        \caption{Actual 2}
        \label{fig:image11}
    \end{subfigure}
    \begin{subfigure}[b]{0.15\textwidth}
        \centering
        \includegraphics[width=\textwidth]{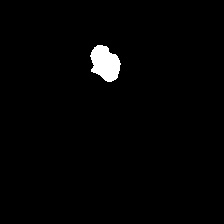}
        \caption{Actual 3}
        \label{fig:image12}
    \end{subfigure}
    \begin{subfigure}[b]{0.15\textwidth}
        \centering
        \includegraphics[width=\textwidth]{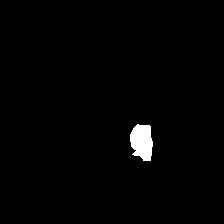}
        \caption{Actual 4}
        \label{fig:image13}
    \end{subfigure}
    \begin{subfigure}[b]{0.15\textwidth}
        \centering
        \includegraphics[width=\textwidth]{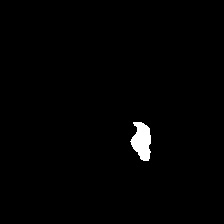}
        \caption{Actual 5}
        \label{fig:image14}
    \end{subfigure}
    \begin{subfigure}[b]{0.15\textwidth}
        \centering
        \includegraphics[width=\textwidth]{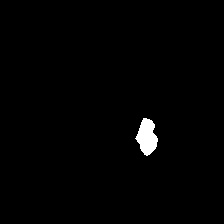}
        \caption{Actual 6}
        \label{fig:image15}
    \end{subfigure}
     
     \vspace{0.15cm} 
     \begin{subfigure}[b]{0.15\textwidth}
        \centering
        \includegraphics[width=\textwidth]{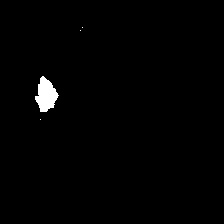}
        \caption{Pred. 1}
        \label{}
    \end{subfigure}
    \begin{subfigure}[b]{0.15\textwidth}
        \centering
        \includegraphics[width=\textwidth]{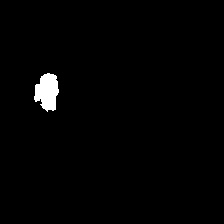}
        \caption{Pred. 2}
         
    \end{subfigure}
    \begin{subfigure}[b]{0.15\textwidth}
        \centering
        \includegraphics[width=\textwidth]{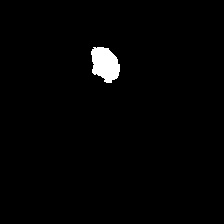}
        \caption{Pred. 3}
         
    \end{subfigure}
    \begin{subfigure}[b]{0.15\textwidth}
        \centering
        \includegraphics[width=\textwidth]{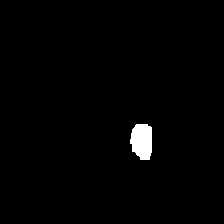}
        \caption{Pred. 4}
         
    \end{subfigure}
    \begin{subfigure}[b]{0.15\textwidth}
        \centering
        \includegraphics[width=\textwidth]{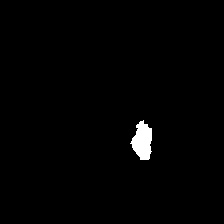}
        \caption{Pred. 5}
        \label{fig:image19}
    \end{subfigure}
    \begin{subfigure}[b]{0.15\textwidth}
        \centering
        \includegraphics[width=\textwidth]{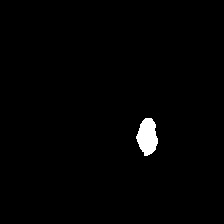}
        \caption{Pred. 6}
        \label{fig:image20}
    \end{subfigure}

     \vspace{0.15cm} 
      \begin{subfigure}[b]{0.15\textwidth}
        \centering
        \includegraphics[width=\textwidth]{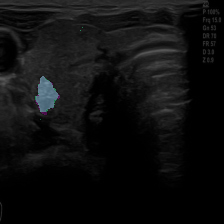}
        \caption{Overlaid 1}
        \label{}
    \end{subfigure}
    \begin{subfigure}[b]{0.15\textwidth}
        \centering
        \includegraphics[width=\textwidth]{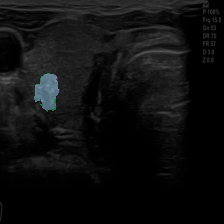}
        \caption{Overlaid 2}
         
    \end{subfigure}
    \begin{subfigure}[b]{0.15\textwidth}
        \centering
        \includegraphics[width=\textwidth]{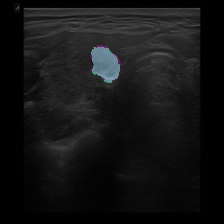}
        \caption{Overlaid 3}
         
    \end{subfigure}
    \begin{subfigure}[b]{0.15\textwidth}
        \centering
        \includegraphics[width=\textwidth]{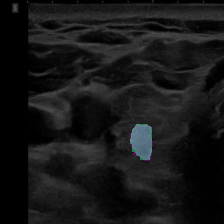}
        \caption{Overlaid 4}
         
    \end{subfigure}
    \begin{subfigure}[b]{0.15\textwidth}
        \centering
        \includegraphics[width=\textwidth]{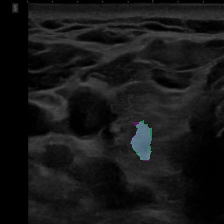}
        \caption{Overlaid 5}
         
    \end{subfigure}
    \begin{subfigure}[b]{0.15\textwidth}
        \centering
        \includegraphics[width=\textwidth]{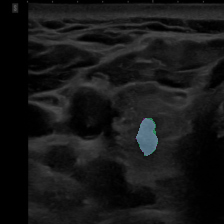}
        \caption{Overlaid 6}
         
    \end{subfigure}
    
    \caption{Evaluating the qualitative performance of DualSwinUnet++. The first and the second rows show six randomly selected test instances along with their associated labels. Third row depicts the models' prediction segmentation for each image, while the last row overlays the first three rows, in which the white, pink, and green areas indicate true positive, false negative, and false positive, respectively. The Feature heatmap is provided in Appendix \ref{feature-heatmap}.}
    \label{f3}
\end{figure}



\begin{figure}[ht]
    \centering
    \begin{subfigure}[b]{0.10\textwidth}
        \centering
        \includegraphics[width=\textwidth]{orig_img1.jpg}
        \caption{Image:1}
         
    \end{subfigure}
    \begin{subfigure}[b]{0.10\textwidth}
        \centering
        \includegraphics[width=\textwidth]{orig_img2.jpg}
        \caption{Image:2}
        \label{fig:image6-2}
    \end{subfigure}
    \begin{subfigure}[b]{0.10\textwidth}
        \centering
        \includegraphics[width=\textwidth]{orig_img3.jpg}
        \caption{Image:3}
        \label{fig:image7-2}
    \end{subfigure}
    \begin{subfigure}[b]{0.10\textwidth}
        \centering
        \includegraphics[width=\textwidth]{orig_img4.jpg}
        \caption{Image:4}
        \label{fig:image8-2}
    \end{subfigure}
    \begin{subfigure}[b]{0.10\textwidth}
        \centering
        \includegraphics[width=\textwidth]{orig_img5.jpg}
        \caption{Image:5}
        \label{fig:image9-2}
    \end{subfigure}
    \begin{subfigure}[b]{0.10\textwidth}
        \centering
        \includegraphics[width=\textwidth]{orig_img6.jpg}
        \caption{Image:6}
        \label{fig:image10-2}
    \end{subfigure}
    \vspace{0.15cm} 
    \\
    \begin{subfigure}[b]{0.10\textwidth}
        \centering
        \includegraphics[width=\textwidth]{true_ptmc_1.jpg}
        \caption{Actual:1}
        \label{}
    \end{subfigure}
    \begin{subfigure}[b]{0.10\textwidth}
        \centering
        \includegraphics[width=\textwidth]{true_ptmc_2.jpg}
        \caption{Actual:2}
        \label{fig:image11-2}
    \end{subfigure}
    \begin{subfigure}[b]{0.10\textwidth}
        \centering
        \includegraphics[width=\textwidth]{true_ptmc_3.jpg}
        \caption{Actual:3}
        \label{fig:image12-2}
    \end{subfigure}
    \begin{subfigure}[b]{0.10\textwidth}
        \centering
        \includegraphics[width=\textwidth]{true_ptmc_4.jpg}
        \caption{Actual:4}
        \label{fig:image13-2}
    \end{subfigure}
    \begin{subfigure}[b]{0.10\textwidth}
        \centering
        \includegraphics[width=\textwidth]{true_ptmc_5.jpg}
        \caption{Actual:5}
        \label{fig:image14-2}
    \end{subfigure}
    \begin{subfigure}[b]{0.10\textwidth}
        \centering
        \includegraphics[width=\textwidth]{true_ptmc_6.jpg}
        \caption{Actual:6}
        \label{fig:image15-2}
    \end{subfigure}
     \\
     \vspace{0.15cm}     
     \begin{subfigure}[b]{0.10\textwidth}
        \centering
        \includegraphics[width=\textwidth]{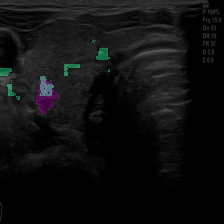}
        \caption{Unet:1}
        \label{}
    \end{subfigure}
    \begin{subfigure}[b]{0.10\textwidth}
        \centering
        \includegraphics[width=\textwidth]{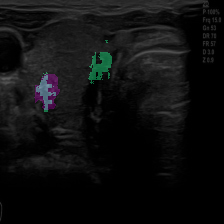}
        \caption{Unet:2}
        \label{fig:image16-2}
    \end{subfigure}
    \begin{subfigure}[b]{0.10\textwidth}
        \centering
        \includegraphics[width=\textwidth]{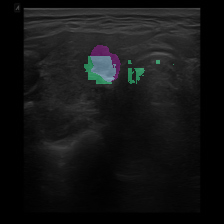}
        \caption{Unet:3}
        \label{fig:image17-2}
    \end{subfigure}
    \begin{subfigure}[b]{0.10\textwidth}
        \centering
        \includegraphics[width=\textwidth]{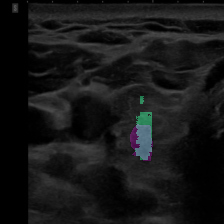}
        \caption{Unet:4}
        \label{fig:image18-2}
    \end{subfigure}
    \begin{subfigure}[b]{0.10\textwidth}
        \centering
        \includegraphics[width=\textwidth]{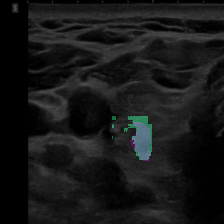}
        \caption{Unet:5}
         
    \end{subfigure}
    \begin{subfigure}[b]{0.10\textwidth}
        \centering
        \includegraphics[width=\textwidth]{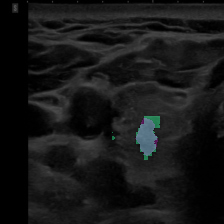}
        \subcaption{Unet:6}
         
    \end{subfigure}
    \\
    
     \vspace{0.15cm} 
      \begin{subfigure}[b]{0.10\textwidth}
        \centering
        \includegraphics[width=\textwidth]{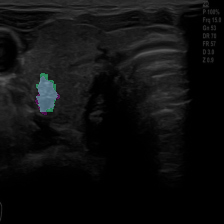}
        \caption{SwinUnet:1}
        \label{}
    \end{subfigure}
    \begin{subfigure}[b]{0.10\textwidth}
        \centering
        \includegraphics[width=\textwidth]{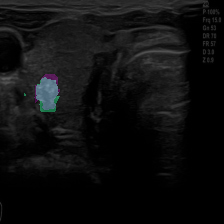}
        \caption{SwinUnet:2}
         
    \end{subfigure}
    \begin{subfigure}[b]{0.10\textwidth}
        \centering
        \includegraphics[width=\textwidth]{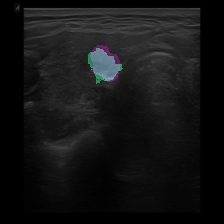}
        \caption{SwinUnet:3}
         
    \end{subfigure}
    \begin{subfigure}[b]{0.10\textwidth}
        \centering
        \includegraphics[width=\textwidth]{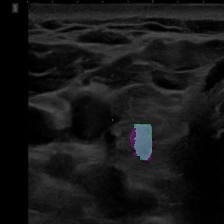}
        \caption{SwinUnet:4}
         
    \end{subfigure}
    \begin{subfigure}[b]{0.10\textwidth}
        \centering
        \includegraphics[width=\textwidth]{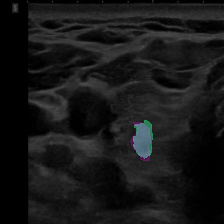}
        \caption{SwinUnet:5}
         
    \end{subfigure}
    \begin{subfigure}[b]{0.10\textwidth}
        \centering
        \includegraphics[width=\textwidth]{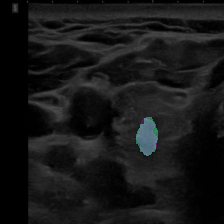}
        \caption{SwinUnet:6}
         
    \end{subfigure}
    \\
  \vspace{0.15cm} 
     \begin{subfigure}[b]{0.10\textwidth}
        \centering
        \includegraphics[width=\textwidth]{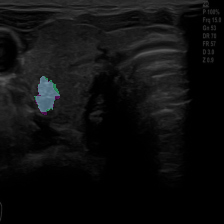}
        \caption{TRFE-Net:1}
        \label{}
    \end{subfigure}
    \begin{subfigure}[b]{0.10\textwidth}
        \centering
        \includegraphics[width=\textwidth]{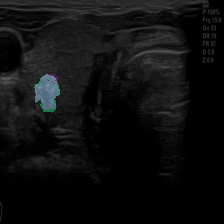}
        \caption{TRFE-Net:2}
         
    \end{subfigure}
    \begin{subfigure}[b]{0.10\textwidth}
        \centering
        \includegraphics[width=\textwidth]{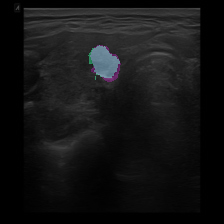}
        \caption{TRFE-Net:3}
         
    \end{subfigure}
    \begin{subfigure}[b]{0.10\textwidth}
        \centering
        \includegraphics[width=\textwidth]{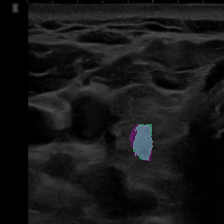}
        \caption{TRFE-Net:4}
         
    \end{subfigure}
    \begin{subfigure}[b]{0.10\textwidth}
        \centering
        \includegraphics[width=\textwidth]{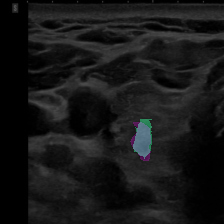}
        \caption{TRFE-Net:5}
        \label{fig:image19-2}
    \end{subfigure}
    \begin{subfigure}[b]{0.10\textwidth}
        \centering
        \includegraphics[width=\textwidth]{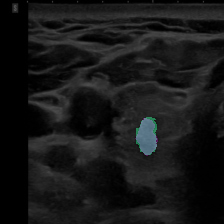}
        \subcaption{TRFE-Net:6}
        \label{fig:image20-2}
    \end{subfigure}
    \\
  \vspace{0.15cm} 
     \begin{subfigure}[b]{0.10\textwidth}
        \centering
        \includegraphics[width=\textwidth]{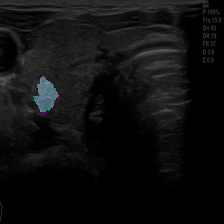}
        \caption{Medsegdiff:1}
        \label{}
    \end{subfigure}
    \begin{subfigure}[b]{0.10\textwidth}
        \centering
        \includegraphics[width=\textwidth]{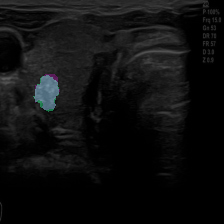}
        \caption{Medsegdiff:2}
         
    \end{subfigure}
    \begin{subfigure}[b]{0.10\textwidth}
        \centering
        \includegraphics[width=\textwidth]{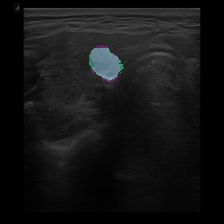}
        \caption{Medsegdiff:3}
         
    \end{subfigure}
    \begin{subfigure}[b]{0.10\textwidth}
        \centering
        \includegraphics[width=\textwidth]{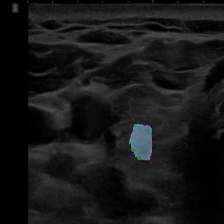}
        \caption{Medsegdiff:4}
         
    \end{subfigure}
    \begin{subfigure}[b]{0.10\textwidth}
        \centering
        \includegraphics[width=\textwidth]{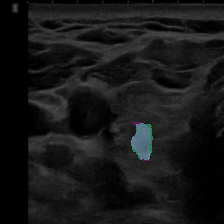}
        \caption{Medsegdiff:5}
         
    \end{subfigure}
    \begin{subfigure}[b]{0.10\textwidth}
        \centering
        \includegraphics[width=\textwidth]{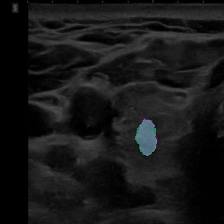}
        \subcaption{Medsegdiff-v2:6}
         
    \end{subfigure}
    \\
     \vspace{0.15cm} 
      \begin{subfigure}[b]{0.10\textwidth}
        \centering
        \includegraphics[width=\textwidth]{combined_ptmc_1Dice_0.969.png}
        \caption{Proposed:1}
        \label{}
    \end{subfigure}
    \begin{subfigure}[b]{0.10\textwidth}
        \centering
        \includegraphics[width=\textwidth]{combined_ptmc_2Dice_0.963.png}
        \caption{Proposed:2}
         
    \end{subfigure}
    \begin{subfigure}[b]{0.10\textwidth}
        \centering
        \includegraphics[width=\textwidth]{combined_ptmc_3Dice_0.955.png}
        \caption{Proposed:3}
         
    \end{subfigure}
    \begin{subfigure}[b]{0.10\textwidth}
        \centering
        \includegraphics[width=\textwidth]{combined_ptmc_4Dice_0.963.png}
        \caption{Proposed:4}
         
    \end{subfigure}
    \begin{subfigure}[b]{0.10\textwidth}
        \centering
        \includegraphics[width=\textwidth]{combined_ptmc_5Dice_0.968.png}
        \caption{Proposed:5}
         
    \end{subfigure}
    \begin{subfigure}[b]{0.10\textwidth}
        \centering
        \includegraphics[width=\textwidth]{combined_ptmc_6Dice_0.967.png}
        \caption{Proposed:6}
         
    \end{subfigure}
    
    \caption{Qualitative comparison of overlaid segmentation results across several related models.}
    \label{f5}
\end{figure}

Similarly to the Jaccard Index, \(A\) and \(B\) represent the ground truth and the predicted segmentation. The Dice Coefficient also ranges from 0 to 1, where 0 means no overlap and 1 signifies a perfect match.

\section{Results}
\subsection{Qualitative Evaluation}
The first qualitative experiment evaluates the segmentation performance of DualSwinUnet++ on six randomly selected test samples from our collected dataset. As shown in Figure~\ref{f3}, the first and second rows display the original ultrasound images and their corresponding ground truth PTMC annotations, respectively. The third row shows the PTMC predictions produced by our model, and the final row overlays ground truth and predicted masks, with white representing true positives, pink for false negatives, and green for false positives. From these visualizations, several key observations can be made:  
(1) DualSwinUnet++ consistently captures the full extent of the PTMC regions without significant under-segmentation, indicating high recall.  
(2) The contours of the predicted regions closely follow the actual PTMC boundaries, even in complex regions with blurred or low-contrast textures.  
(3) The presence of false positives and false negatives is minimal across all samples, which reflects the model’s capacity to distinguish PTMC from surrounding tissues, despite challenges such as acoustic shadows or tissue heterogeneity.

To further compare model performance qualitatively, Figure~\ref{f5} presents an overlay comparison of segmentation predictions from several state-of-the-art models, including UNet, Swin-Unet, TRFE-Net, Medsegdiff-v2, and our proposed method. It can be seen that:

- UNet and UNet++ often produce incomplete segmentations, missing portions of the tumor or introducing boundary noise, particularly in low-contrast areas.\\
- Swin-Unet and TRFE-Net show improved precision but occasionally exhibit over-segmentation into non-tumorous regions, especially in elongated thyroid structures.\\
- Medsegdiff-v2 demonstrates smoother contours but fails to capture fine structural boundaries in several examples.

In contrast, DualSwinUnet++ exhibits the most consistent alignment with the ground truth across all six samples. Notably, it avoids both common error types: over-segmentation into the background and under-segmentation of internal tumor areas. This improvement can be attributed to the model's architectural features, such as its thyroid-aware conditioning via the first decoder and residual information flow between decoders. Overall, the qualitative results reinforce the numerical superiority reported in Table~\ref{table:1} and Table~\ref{table:confusion_metrics}, showcasing not only higher aggregate metrics but also consistently better localization and boundary precision on individual cases.

\subsection{Quantitative Evaluation}
Table~\ref{table:1} reports the segmentation accuracy of all models in terms of Jaccard Index and Dice Coefficient on both the collected clinical dataset and the TN3K benchmark. Across both datasets, DualSwinUnet++ achieves the highest scores, indicating its superior ability to correctly identify the tumor region while minimizing both false positives and false negatives.

On our collected dataset, DualSwinUnet++ achieves a Dice score of 84.36\% and a Jaccard index of 71.23\%, outperforming the next-best model (TRFE-Net) by 0.75\% in Dice and over 2\% in Jaccard. This improvement is particularly meaningful considering the small size and ambiguous boundaries of PTMC lesions, which often result in segmentation drift. The Jaccard metric, being more sensitive to over- and under-segmentation, highlights the model's precise spatial localization.

On the TN3K dataset, which includes a broader variety of thyroid nodule appearances, DualSwinUnet++ also maintains its advantage, with a Dice of 81.24\% and a Jaccard of 69.14\%. Notably, its performance margin over the Transformer-based Swin-Unet++ and TRFE-Net remains consistent, which speaks to the generalizability of the architecture.

The improvements can be attributed to two core architectural features of DualSwinUnet++: (1) the use of thyroid-gland-aware spatial guidance from the first decoder, and (2) the residual information fusion between decoders that enhances discrimination between tumor and non-tumor regions. These results demonstrate that our model is not only accurate in segmenting PTMC but also robust across datasets with varying imaging conditions and tumor presentations.

\begin{table}[h!]
\centering
\caption{Model performance comparison in terms of Dice and Jaccard scores. (Left) Results on our collected dataset. (Right) Results on the TN3K dataset. Each experiment is repeated three times and the average results are reported. }
\begin{tabular}{cc}
\begin{tabular}{|>{\centering\arraybackslash}m{3cm}|>{\centering\arraybackslash}m{2cm}|>{\centering\arraybackslash}m{1.5cm}|}
\hline
\textbf{Model} & \textbf{Jaccard(\%)} & \textbf{Dice(\%)} \\ \hline
Unet & 63.17 &77.41  \\ \hline
Unet++ & 64.42 &78.60  \\ \hline
Swin-Unet & 65.65 &79.87  \\ \hline
Swint-Unet++ & 67.20 &82.15  \\ \hline
TRFE-Net &69.19  &83.61  \\ \hline
\textbf{DualSwinUnet++} & \textbf{71.23} &\textbf{84.36}  \\ \hline

\end{tabular}
&
\begin{tabular}{|>{\centering\arraybackslash}m{3cm}|>{\centering\arraybackslash}m{2cm}|>{\centering\arraybackslash}m{1.5cm}|}
\hline
\textbf{Model} & \textbf{Jaccard(\%)} & \textbf{Dice(\%)} \\ \hline
Unet & 62.85 &76.71  \\ \hline
Unet ++ & 63.41 &77.69  \\ \hline
Swin-Unet & 65.24 &79.39  \\ \hline
Swint-Unet++ & 66.32 &80.10  \\ \hline
TRFE-Net &67.65  &79.14   \\ \hline
\textbf{DualSwinUnet++} & \textbf{69.14} & \textbf{81.24} \\ \hline

\end{tabular}
\end{tabular}

\label{table:1}
\end{table}
In addition to Dice and Jaccard metrics, we further analyzed segmentation performance using true positive (TP), false positive (FP), false negative (FN), and F1-score metrics, as presented in Table~\ref{table:confusion_metrics}. These metrics provide a more detailed view of the model’s precision-recall trade-off and its behavior in different error modes—such as missed detections and over-segmentation.

On our collected dataset, DualSwinUnet++ achieves the highest true positive rate (71.5\%) while maintaining the lowest false positive rate (14.2\%), indicating that it not only captures the tumor regions effectively but also avoids misclassifying non-tumor regions. Compared to TRFE-Net and Swint-Unet++, our model reduces false positives by over 1.5\% and improves F1-score by approximately 0.75\%. It also maintains a balanced FN rate (14.3\%)—lower than Unet variants and comparable to TRFE-Net—demonstrating a strong ability to avoid under-segmentation.

On the TN3K dataset, DualSwinUnet++ continues to outperform other models, achieving an F1-score of 81.24\%. It shows the best combination of low FP (15.3\%) and FN (15.2\%) rates across all tested methods. This consistency in detection accuracy across datasets reinforces the model’s robustness to inter-patient variability and imaging noise.

The observed improvements in F1-score directly reflect the contributions of our architecture: specifically, the conditioning of the second decoder on thyroid gland features from the first decoder improves localization precision, while the independent linear projections ensure stable gradient flows that mitigate overfitting and segmentation drift. These results confirm that DualSwinUnet++ is not only accurate in terms of overlap metrics but also excels in avoiding common segmentation pitfalls across varied clinical cases.

\begin{table}[h!]
\centering
\caption{Comparison of model performance using true positives (TP), false positives (FP), false negatives (FN), and F1-score on both datasets.}
\label{table:confusion_metrics}
\begin{subtable}{\textwidth}
\centering
\caption{Results on our collected dataset.}
\begin{tabular}{|l|c|c|c|c|}
\hline
\textbf{Model} & \textbf{TP (\%)} & \textbf{FP (\%)} & \textbf{FN (\%)} & \textbf{F1-score (\%)} \\ \hline
Unet & 63.2 & 18.9 & 17.9 & 77.41 \\ \hline
Unet++ & 64.4 & 17.8 & 17.8 & 78.60 \\ \hline
Swin-Unet & 66.3 & 22.8 & \textbf{10.9} & 79.87 \\ \hline
Swint-Unet++ & 67.4 & 16.8 & 15.8 & 82.15 \\ \hline
TRFE-Net & 69.2 & 15.9 & 14.9 & 83.61 \\ \hline
\textbf{DualSwinUnet++} & \textbf{71.5} & \textbf{14.2} & 14.3 & \textbf{84.36} \\ \hline
\end{tabular}
\end{subtable}

\vspace{0.5cm}

\begin{subtable}{\textwidth}
\centering
\caption{Results on the TN3K dataset.}
\begin{tabular}{|l|c|c|c|c|}
\hline
\textbf{Model} & \textbf{TP (\%)} & \textbf{FP (\%)} & \textbf{FN (\%)} & \textbf{F1-score (\%)} \\ \hline
Unet & 62.2 & 18.5 & 19.3 & 76.71 \\ \hline
Unet++ & 63.6 & 18.4 & 18.0 & 77.69 \\ \hline
Swin-Unet & 65.5 & 17.3 & 17.2 & 79.39 \\ \hline
Swint-Unet++ & 66.6 & 16.7 & 16.7 & 80.10 \\ \hline
TRFE-Net & 67.9 & 16.0 & 16.1 & 79.14 \\ \hline
\textbf{DualSwinUnet++} & \textbf{69.5} & \textbf{15.3} & \textbf{15.2} & \textbf{81.24} \\ \hline
\end{tabular}
\end{subtable}
\end{table}

\begin{figure}[h!]
    \centering
    \includegraphics[width=0.8\textwidth]{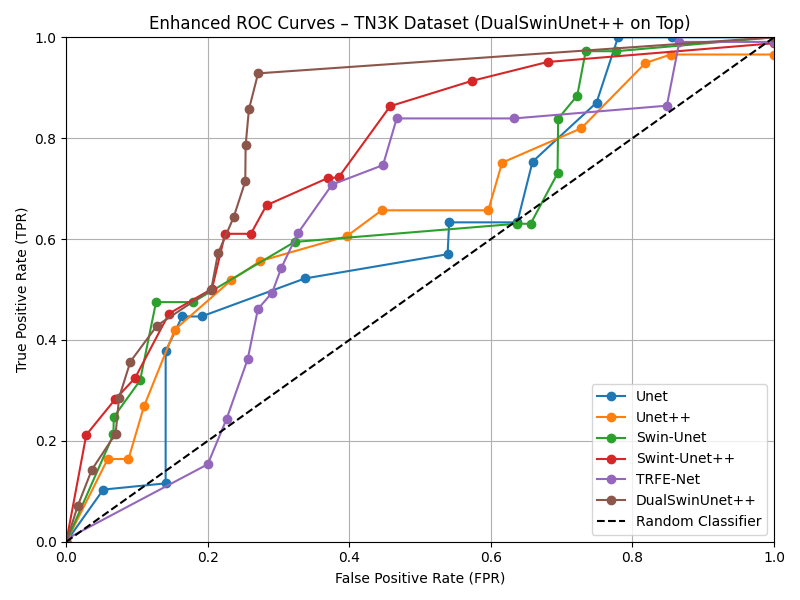}  
    \caption{ROC curves of all evaluated models on the TN3K dataset. The curves plot the True Positive Rate (TPR) against the False Positive Rate (FPR) for varying threshold values. DualSwinUnet++ consistently achieves the highest TPR across all FPR levels, indicating superior sensitivity and segmentation discrimination. The dashed diagonal line represents the performance of a random classifier.}
    \label{fig:roc_curve}
\end{figure}

To provide additional visual evidence of model discrimination capability, we include the ROC (Receiver Operating Characteristic) curves for all evaluated models on the TN3K dataset, as shown in Figure~\ref{fig:roc_curve}. The ROC curve plots the True Positive Rate (TPR) against the False Positive Rate (FPR) across various threshold levels. As illustrated, \textbf{DualSwinUnet++} consistently dominates other models across the entire curve, achieving the highest TPR for nearly all FPR values. This suggests superior sensitivity and robustness in differentiating PTMC regions from background tissue. In contrast, earlier models like UNet and Unet++ exhibit limited separability, with curves remaining closer to the diagonal (random classifier), especially at lower thresholds. These results visually reinforce the quantitative metrics reported earlier and highlight DualSwinUnet++’s strength in clinical decision contexts where detection sensitivity is critical.

\subsection{Speed and Efficiency Comparison}
The second experiment evaluates the performance of DualSwinUnet++ in both segmentation accuracy and prediction time per sample, comparing it with several state-of-the-art models, including nnUnet, Swin-UNetr, Unetr, SegDiff, and Medsegdiff-v2. As shown in Table \ref{t2}, DualSwinUnet++ achieves the highest Dice score of 83.36\%, outperforming all other models in terms of segmentation accuracy. This result demonstrates the effectiveness of our model in leveraging thyroid gland information and utilizing dual decoders to enhance the PTMC segmentation. When it comes to prediction time, DualSwinUnet++ provides a balance between accuracy and efficiency. With an average prediction time of 0.18 seconds per sample, it is slightly slower than nnUnet and Swin-UNetr but significantly faster thanUnetr, SegDiff, and Medsegdiff-v2. This result indicates that DualSwinUnet++ achieves superior segmentation performance without a significant trade-off in computational efficiency. Moreover, the prediction time of less than 0.2 seconds per sample highlights the potential of DualSwinUnet++ model for online prediction tasks. This rapid inference capability makes it promising for use as an AI assistant during RFA surgery, where real-time decision support is critical (Appendix \ref{justification}). 

\begin{table}[h]
\centering
\caption{Comparison of models based on Dice scores on our collected dataset, along with the average prediction time per sample.}
\begin{tabular}{|l|c|c|}
\hline
\textbf{Model}             & \textbf{Dice(\%)} & \textbf{Average Prediction Time (seconds)} \\ \hline
nnUnet                     & 80.35                        & \textbf{0.14} second              \\ \hline
Swin-UNetr                 & 81.16                        & 0.15 second              \\ \hline
UNetr                      & 79.68                        & 0.25 second              \\ \hline
SegDiff                    & 80.87                        & 0.36 second              \\ \hline
Medsegdiff-v2                    & 80.68                        & 0.31 second              \\ \hline
\textbf{DualSwinUnet++}   & \textbf{83.36}               & 0.18 second     \\ \hline
\end{tabular}
\label{t2}
\end{table}

\subsection{Ablation Study }
This subsection explores the impact of different components of DualSwinUnet++ on its segmentation performances. Specifically, we evaluate the effect of the first decoder, the types of skip connections and the number of skip connections on the model’s overall performance.

\textbf{Impact of the first decoder:}
To evaluate the effectiveness of our proposed architecture, we assess the the effect of adding thyroid feature in DualSwinUnet++. Specifically, we aim to understand how the first decoder enhances  PTMC segmentation through the second decoder. To verify this, we removed the first decoder from the architecture and repeated the experiments outlined in Table \ref{table:1}. This change in the network downgrades the model to vanilla Swin-Unet. Interestingly, we observed a performance drop of 4.98\% on our dataset and 3.11\%  on the TN3K benchmark. These results suggest that the inclusion of thyroid gland boundaries significantly enhances the model’s performance.

\textbf{Impact of skip connections types:}
DualSwinUnet++ employs Unet-like skip connections, where residual features are concatenated with current features and passed through a linear layer. This experiment explores the impact of using additive skip connections instead of concatenated ones. In this variation,  residual features are added to the current features before passing them to the Swin-Transformers blocks. This modifications resulted in Dice score drops of 6.51\% on our dataset and 5.11\%  on the TN3K benchmark. Hence, we conclude that the choice of Unet-like concatenated skip connections outperforms additive skip connections in this context.   

\textbf{Impact of the number of skip connections:}
DualSwinUnet++ employs six skip connections, with two per each layer. This experiment evaluates the impact of varying the number of skip connections on the performance. Similar to Swin-Unet, we use three skip connections from the encoder to the first decoder and three additional connections from the first decoder to the second one. In this experiment, we vary the number of skip connections from $0$ to $6$ and report the results in Table \ref{t3}. The results indicate that increasing the number of skip connections consistently improves the model's performance, as evidenced by higher Jaccard and Dice scores on both our dataset and the TN3K benchmark. The highest performance is achieved with 6 skip connections.

\begin{table}
	
	\begin{center}
		\caption{Impact of the number of skip connections on the performance of the DualSwinUnet++ model. The table presents the results as the number of skip connections varies from 0 to 6, demonstrating how these changes affect the model's performance.}
		\label{t3}
		\resizebox{0.5\columnwidth}{!}{%
			\begin{tabular}{lcccc}
				\toprule
				& \multicolumn{2}{c}{ \textbf{Our Dataset}} & \multicolumn{2}{c}{ \textbf{TN3K}} \\\cmidrule(lr){2-3} \cmidrule(lr){4-5}
				& Jaccard &Dice &Jaccard &Dice\\ \midrule
				\textbf{Skip Connections}\\
				\;0 &65.21 &78.17 &62.34 &74.69 \\ 
				\;1 &67.14 &79.34 &64.24 &75.84  \\ 
				\;2 &68.23 &80.53 &63.72 &77.12   \\  
				\;3 &69.47 &81.72 &65.16 &78.46   \\ 
                \;4 &70.59 &82.46 &66.84 &79.13   \\ 
                \;5 &71.08 &83.06 &68.31 &80.69   \\ 
                \;6 &\textbf{71.23} &\textbf{84.36} &\textbf{69.14} &\textbf{81.24}   \\
				\bottomrule
			\end{tabular}
		}
	\end{center}
\end{table}

\section{Discussion}

The proposed DualSwinUnet++ model demonstrates strong potential for enhancing PTMC segmentation under ultrasound imaging by leveraging thyroid-aware contextual encoding and a dual-decoder architecture. As supported by both quantitative metrics and visual evidence, our model outperforms a wide range of state-of-the-art methods in segmentation accuracy, recall, and boundary localization, while maintaining a competitive inference time suitable for real-time deployment.

The ablation studies reinforce the effectiveness of our architectural innovations. Removing the first decoder, which provides thyroid-region guidance, results in a significant drop in Dice and Jaccard scores—highlighting its crucial role in localizing tumors within the anatomical context. Similarly, varying the type and number of skip connections confirmed that our adopted Unet-style residual fusion and complete skip pathway design offer substantial accuracy gains over minimal or additive alternatives. These results illustrate that the performance of DualSwinUnet++ is tightly linked to its design choices, and its strength lies in the tailored integration of spatial priors with efficient feature fusion.

\textbf{Limitations:} Despite the strong performance, our model has several practical limitations. First, although we designed DualSwinUnet++ to handle noise and structural variability in ultrasound images, its robustness may degrade under extreme acoustic shadowing or image artifacts caused by microbubbles during ablation—conditions that were not explicitly annotated or stratified in our dataset. Second, while our quantitative comparisons and ROC curves show improved generalization on the public TN3K dataset, these datasets may still lack sufficient demographic and institutional diversity. For instance, all training data were collected from a single clinical source with similar acquisition protocols, and the model has not yet been validated across different ultrasound machines or healthcare centers.

Another performance-related limitation is the model's reliance on a relatively high-quality initial thyroid gland segmentation. In rare cases where the first decoder inaccurately segments the thyroid region—due to anatomical anomalies or severe imaging noise—the second decoder’s PTMC segmentation may also drift. A potential mitigation strategy would be to incorporate uncertainty modeling or confidence-based weighting between decoder pathways.

\textbf{Future Direction:} Future research will address these limitations by collecting and evaluating the model on a more diverse and larger-scale dataset sourced from multiple hospitals and ultrasound devices. We also plan to extend the model to handle live video-based ultrasound sequences to support continuous real-time inference during RFA procedures. Moreover, adding self-supervised pretraining or domain-adaptive mechanisms may further improve performance under challenging imaging conditions or unseen patient populations. Finally, integrating confidence estimation and active learning can enable clinicians to identify cases needing review and facilitate iterative model improvement through clinical feedback.

\section{Conclusion}
\label{sec:Conclusion}
This paper proposed DualSwinUnet++, a novel model that incorporates thyroid gland information to enhance PTMC segmentation accuracy. The model leverages a dual decoder architectures, with the first decoder providing thyroid features to the second, leading to significant improvements in segmentation performance. Evaluation results demonstrate that DualSwinUnet++ outperforms existing models. Ablation studies further confirm the importance of incorporating thyroid features, as their removal resulted in substantial performance drops, validating our design choice. Furthermore, time assessment results indicate that DualSwinUnet++ also holds potential for real-time applications in clinical settings, such as AI-assisted decision-making during RFA surgeries.

\bigskip 
\noindent 
\textbf{Ethics Statement}
All procedures performed in this study were in compliance with relevant laws and institutional guidelines. Consent was obtained from all human subjects prior to their participation in the study, in accordance with ethical standards. The privacy and confidentiality of all participants were strictly maintained throughout the research process.

\bibliographystyle{unsrt}  
\bibliography{references}  
\clearpage
\textbf{APPENDIX}
\appendix

\section{Dataset Acquisition and Annotation Details}

The first dataset used in this study consists of 691 ultrasound images collected from 101 anonymous patients who underwent radiofrequency ablation (RFA) procedures. These images were acquired using clinical-grade B-mode ultrasound imaging systems equipped with high-frequency transducers typically used for thyroid imaging.

\textbf{Acquisition Devices and Settings}:
\begin{itemize}
    \item \textbf{Ultrasound Systems}: Philips Affiniti 70, Samsung RS80A
    \item \textbf{Imaging Mode}: B-mode (brightness mode)
    \item \textbf{Frequency Range}: 15--18 MHz
    \item \textbf{Image Resolution}: 786 $\times$ 531 pixels
    \item \textbf{Scan Settings}: Gain, depth, and focus settings were adjusted per patient anatomy
\end{itemize}

\textbf{Patient Demographics}:
\begin{itemize}
    \item \textbf{Gender Ratio}: 50\% male, 50\% female
    \item \textbf{Age Range}: 25--72 years
\end{itemize}

\textbf{Annotation Process}:
\begin{itemize}
    \item All images were manually annotated by two certified radiologists, each with over 5 years of experience in thyroid imaging.
    \item Annotations included both thyroid gland boundaries and PTMC tumor margins.
    \item Disagreements were resolved through joint discussion, and final annotations were validated by a third expert.
\end{itemize}

Full details of the dataset, including acquisition protocols and annotation guidelines, will be made publicly available through the GitHub repository associated with this paper upon acceptance.

\section{Independent Evaluation of Decoder 1 (Thyroid Segmentation)}

Although only the output of the second decoder (PTMC segmentation) is used at inference time, we evaluated the first decoder separately to assess how well it identifies the thyroid gland, which provides semantic priors to improve PTMC segmentation accuracy.

To perform this analysis, we extracted the output of the first decoder alone and compared it against the ground truth thyroid gland masks using the same evaluation metrics as for PTMC (Jaccard Index and Dice Coefficient). The evaluation was conducted on both our collected dataset and the TN3K benchmark.

\begin{table}[h!]
\centering
\caption{Performance of the First Decoder for Thyroid Gland Segmentation}
\label{tab:first-decoder-eval}
\begin{tabular}{|l|c|c|}
\hline
\textbf{Dataset} & \textbf{Jaccard Index (\%)} & \textbf{Dice Coefficient (\%)} \\
\hline
Our Dataset & 92.45 & 95.34 \\
TN3K Dataset & 91.34 & 94.97 \\
\hline
\end{tabular}
\end{table}

These results confirm that the first decoder performs highly accurate thyroid segmentation, thereby justifying its inclusion in the model and its role in guiding the second decoder.

\section{On The Realtime Justification}
\label{justification}
In clinical settings, particularly during ultrasound-guided procedures such as RFA, the frame rate of typical ultrasound imaging systems ranges between 15 and 30 frames per second \cite{philips_ultrasound}. Although imaging devices acquire frames approximately every 33–66 milliseconds, the clinical decision-making process during RFA surgery typically operates on much slower timescales, where actions such as needle placement adjustments occur over several seconds. Hence, AI inference speeds are expected to support decision-making intervals rather than match imaging frame rates. Our model’s average prediction time of 0.18 seconds per sample ensures updated guidance more than five times per second, which is considerably faster than the typical physician's decision-making pace during RFA interventions \cite{baek2020rfa}. Therefore, in this clinical context, DualSwinUnet++ satisfies the practical requirements of real-time performance, enabling efficient integration into live surgical workflows.

\section{Feature Heatmap}
\label{feature-heatmap}
\begin{figure}[h!]
    \centering
    \includegraphics[width=0.6\textwidth]{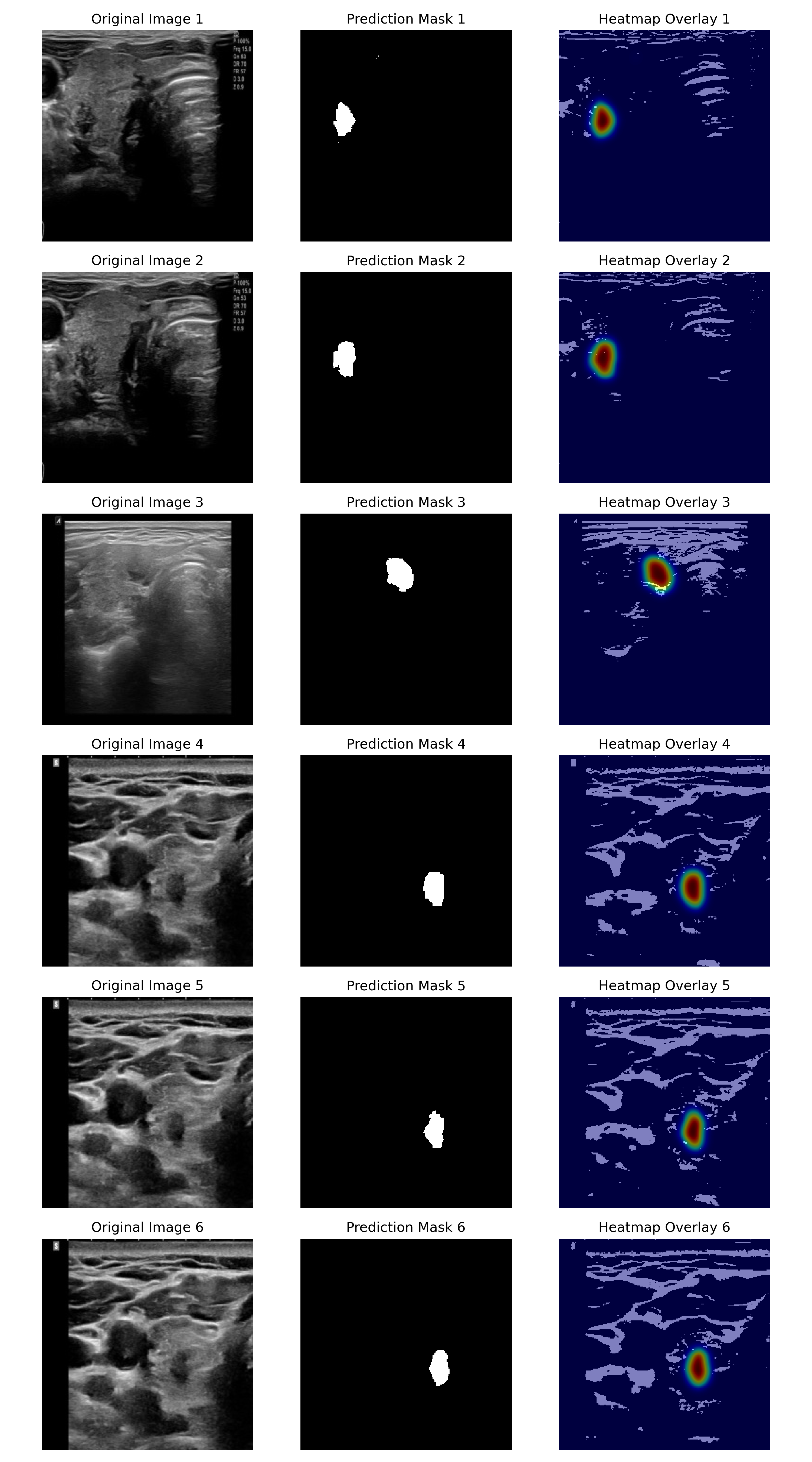} 
    \caption{
    Visualization of feature heatmaps for six representative test samples. Each row displays the original ultrasound image, the predicted segmentation mask, and a heatmap overlay derived from Gaussian-filtered activations of the predicted PTMC region. The heatmap highlights regions where the model places highest confidence during segmentation. As shown, these areas consistently align with the true tumor boundaries, reinforcing the spatial precision and interpretability of the proposed DualSwinUnet++ model.
    }
    \label{fig:feature_heatmap}
\end{figure}

To provide additional spatial insight into the model's focus during prediction, we generated simulated feature heatmaps based on the predicted segmentation masks. As shown in Figure~\ref{fig:feature_heatmap}, each row corresponds to a test case, displaying the original ultrasound image, the predicted binary mask, and a heatmap overlay. The heatmaps were derived using Gaussian-filtered activations over the predicted tumor region, mimicking the spatial attention of a deep model. The overlaid heatmaps highlight the regions where the segmentation model places its highest confidence, which consistently aligns with the tumor boundaries across all cases. This visualization supports the interpretability of our model by showing that the predicted PTMC regions are not only accurate in shape but also localized in semantically relevant anatomical areas.

\section{Hyperparameters}
Table below provides a list of hyper-parameters used in the experiments.

\begin{table}[ht]
    \caption{Hyper-parameters.}
    \centering
    \begin{tabular}{l c}
    \hline\hline 
     Parameter & Value \\ 
    \hline
    BATCH SIZE & $32$ \\
    IMG SIZE & $224$ \\
    DROP RATE & $0.0$ \\
    LABEL SMOOTHING & $0.1$ \\
    PATCH SIZE & $4$ \\
    EMBED DIM & $96$ \\
    DEPTHS & $[2, 2, 6, 2]$ \\
    DECODER DEPTHS & $[2, 2, 6, 2]$ \\
    NUM HEADS & $[3, 6, 12, 24]$ \\
    WINDOW SIZE & $7$ \\
    MLP RATIO & $4.0$ \\
    EPOCHS & $300$ \\
    WARMUP EPOCHS & $20$ \\
    WEIGHT DECAY & $0.05$ \\
    CLIP GRAD & $5.0$ \\
    ACCUMULATION STEPS & $0$ \\
    DECAY EPOCHS & $30$ \\
    DECAY RATE & $0.1$ \\
    BETAS & $(0.9, 0.999)$ \\
    MOMENTUM & $0.9$ \\
    COLOR JITTER & $0.4$ \\
    REPROB & $0.25$ \\
    RECOUNT & $1.0$ \\
    MIXUP & $0.8$ \\
    CUTMIX & $1.0$ \\
    MIXUP PROB & $1.0$ \\
    MIXUP SWITCH\_PROB & $0.5$ \\
    SEED & $0$ \\
    \hline
    \end{tabular}
    \label{t5}
\end{table}

\end{document}